\documentclass[prb,twocolumn,groupaddress,superscriptaddress,showpacs]{revtex4}

\usepackage{graphicx}
\usepackage{amssymb}
\usepackage[usenames]{color}
\usepackage{enumerate}
\usepackage{gensymb}
\usepackage{amsmath}
\usepackage{nicefrac}
\usepackage{multirow}
\usepackage{epstopdf}
\usepackage{float}  
\usepackage{hyperref}

\begin{document}

\title{Common Glass-Forming Spin-Liquid State 
\\in the Pyrochlore Magnets Dy$_2$Ti$_2$O$_7$ and Ho$_2$Ti$_2$O$_7$}

\author{Azar B. Eyvazov} %or A. B.
\thanks{equal contribution} 
\affiliation{LASSP, Department of Physics, Cornell University, Ithaca, NY 14853, USA}

\author{Ritika Dusad}
\thanks{equal contribution} 
\affiliation{LASSP, Department of Physics, Cornell University, Ithaca, NY 14853, USA}

\author{Timothy J. S. Munsie} 
\affiliation{Brockhouse Inst. for Materials Research, McMaster University, Hamilton, ON, Canada}
\affiliation{Department of Physics, McMaster University, Hamilton, Ontario, L8S 4M1, Canada}

\author{Hanna A. Dabkowska} 
\affiliation{Brockhouse Inst. for Materials Research, McMaster University, Hamilton, ON, Canada}

\author{Graeme M. Luke} 
\affiliation{Brockhouse Inst. for Materials Research, McMaster University, Hamilton, ON, Canada}
\affiliation{Department of Physics, McMaster University, Hamilton, Ontario, L8S 4M1, Canada}
\affiliation{Canadian Institute for Advanced Research, Toronto, Ontario, M5G 1Z8, Canada}

\author{Ethan R. Kassner} 
\affiliation{LASSP, Department of Physics, Cornell University, Ithaca, NY 14853, USA}
\affiliation{Current address: Honeywell International Inc., Golden Valley, MN 55422, USA}

\author{J.C. Seamus Davis} 
\affiliation{LASSP, Department of Physics, Cornell University, Ithaca, NY 14853, USA}
\affiliation{CMPMS Department, Brookhaven National Laboratory, Upton, NY 11973, USA}
\affiliation{Tyndall National Institute, University College Cork, Cork T12R5C, Ireland}

\author{Anna Eyal}
\email{anna.eyal@gmail.com} 
\affiliation{LASSP, Department of Physics, Cornell University, Ithaca, NY 14853, USA}

\date{\today}

\begin{abstract}
Despite a well-ordered pyrochlore crystal structure and strong magnetic interactions between the Dy$^{3+}$ or Ho$^{3+}$ ions, no long range magnetic order has been detected in the pyrochlore titanates  Ho$_2$Ti$_2$O$_7$ and Dy$_2$Ti$_2$O$_7$. To explore the actual magnetic phase formed by cooling these materials, we measure their magnetization dynamics using toroidal, boundary-free magnetization transport techniques. We find that the dynamical magnetic susceptibility of both compounds has the same distinctive phenomenology, that is indistinguishable in form from that of the dielectric permittivity of dipolar glass-forming liquids. Moreover, Ho$_2$Ti$_2$O$_7$ and Dy$_2$Ti$_2$O$_7$ both exhibit microscopic magnetic relaxation times that increase along the super-Arrhenius trajectories analogous to those observed in glass-forming dipolar liquids. Thus, upon cooling below about 2K, Dy$_2$Ti$_2$O$_7$ and Ho$_2$Ti$_2$O$_7$ both appear to enter the same magnetic state exhibiting the characteristics of a glass-forming spin-liquid.

\end{abstract}


\maketitle

	\section{Introduction} 

\begin{figure*}[]
\includegraphics[width=1.15\textwidth]{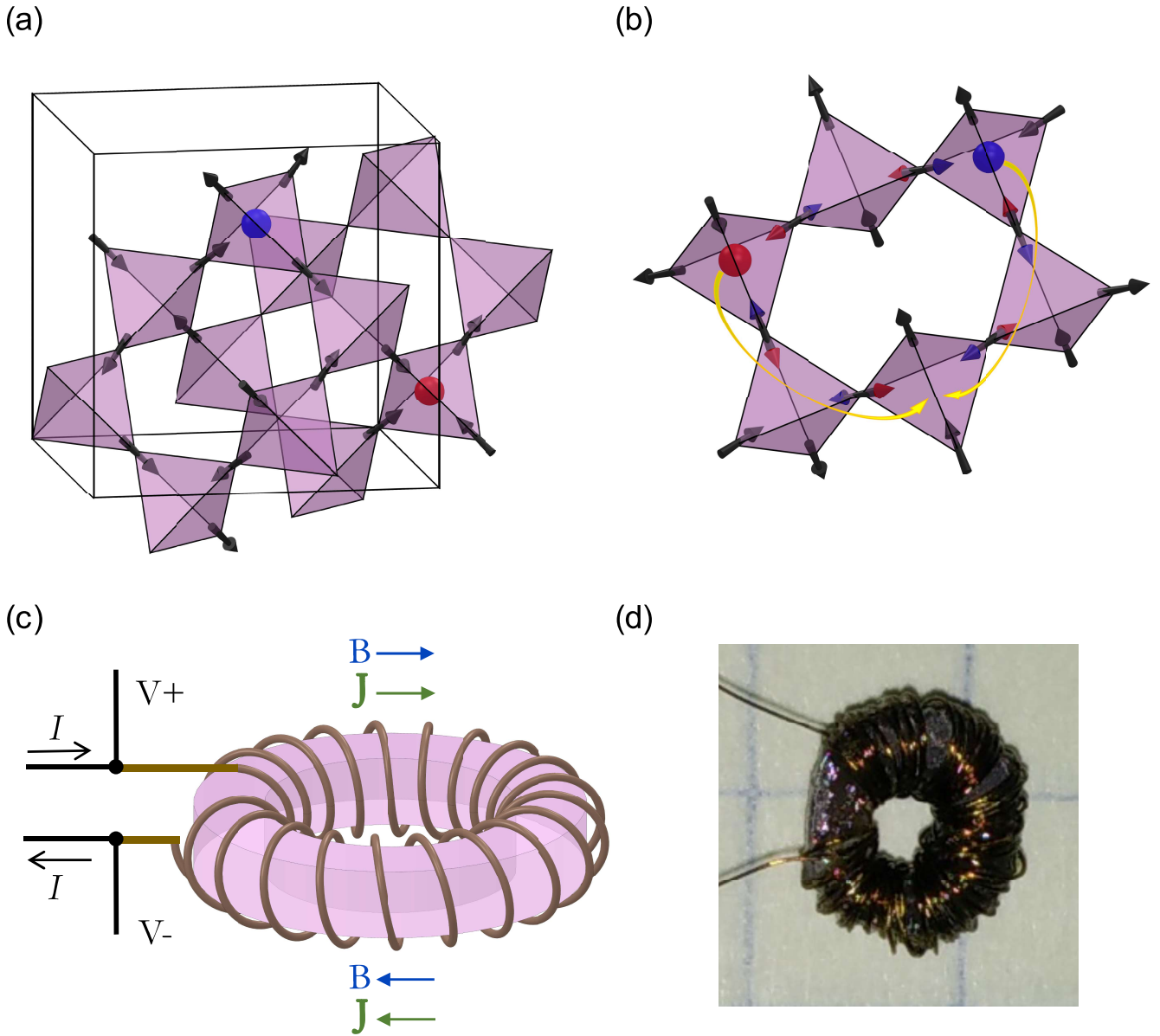} {}
\caption{
An illustration of the spin-ice sample and our measurement setup. 
(a)	Crystal structure of the sublattice of rare earth ions in the pyrochlore titanates. The angular momenta of the ions (Dy$^{3+}$ or Ho$^{3+}$), at the corners of a network of corner sharing tetrahedra, are constrained to point towards or away from the centers of their surrounding tetrahedra (black arrows). The spin-ice ground state then corresponds to a 2-in 2-out spin configuration for each tetrahedron (as can be seen in a tetrahedron on the left).The excitations from this state are magnetic monopole-like quasi-particles. A 3-in 1-out spin state in a tetrahedron can be represented as a monopole (red ball), and a 3-out 1-in state as an anti-monopole (blue ball).
(b)	These monopoles are highly correlated and can be created, transport (yellow arrows) and then get annihilated on loops of varying sizes. The shortest such loop is shown. 
(c)	Measuring the magnetization dynamics in a boundary-free environment is done by applying an azimuthal magnetic field in a torus shaped sample. The Superconducting Toroidal Solenoid (STS) used enables us to eliminate the contribution of the demagnetization effect and to allow for accurate measurements both in time and frequency domains.
(d)	A single crystal toroidal sample of Ho$_2$Ti$_2$O$_7$ wired with a STS of fine NbTi wire. The outer diameter of the sample is 6mm.
}
\label{fig:Fig1}
\end{figure*}

In the pyrochlore lanthanide-oxides with chemical formula A$_2$B$_2$O$_7$, the magnetic rare earth A ions are located at corner sharing tetrahedral sites as shown in Fig. \ref{fig:Fig1}a \cite{Farmer}. These materials support a multitude of exotic magnetic states \cite{Gardner,Lacroix,Castelnovo2008}, such as spin-ice\cite{Ramirez,Snyder}, spin-slush \cite{Rau}, and various candidates for quantum spin-liquids \cite{Balents}. Dy$_2$Ti$_2$O$_7$ and Ho$_2$Ti$_2$O$_7$ have attracted much interest since the discovery of an entropy deficit in both compounds \cite{Ramirez,Lau}, along with a similarity to the thermodynamic properties of water ice. In both Dy$_2$Ti$_2$O$_7$ and Ho$_2$Ti$_2$O$_7$ the large rare-earth magnetic moments of  $\mu\sim$10$\mu_B$ \cite{Balakrishnan}, at the corners of the corner sharing tetrahedral structure, are subject to strong crystal field interactions \cite{Rosenkranz,Jana} so that, in the single-ion ground-state, spins are only allowed to point towards (away) from the centers of the tetrahedra that share it (see Fig. \ref{fig:Fig1}a) \cite{Ramirez}. Under these circumstances, long-range magnetic dipolar interactions are significant when compared to the nearest-neighbor magnetic exchange coupling. The resulting interaction favors a state in which for each tetrahedron two spins point in and two out, that, by analogy to water ice, was dubbed "spin-ice"\cite{Harris}. One might expect such a constraint to result in an ordered magnetic state that is unique \cite{Melko}, but no such state has ever been observed in these materials at zero magnetic field \cite{Fukazawa}.

For both Ho$_2$Ti$_2$O$_7$ and Dy$_2$Ti$_2$O$_7$ the strength of the nearest neighbor dipolar interaction is D$_{nn}\approx2.35$K, while the nearest neighbor exchange interactions are J$_{nn}\approx-0.52$K for Ho$_2$Ti$_2$O$_7$ and J$_{nn}\approx-1.24$K for Dy$_2$Ti$_2$O$_7$ \cite{denHertog,JaubertHoldswort}. Raman spectroscopy reveals that the phonon spectra of these two materials differ only slightly and that the crystal field parameters are very similar \cite{Lummen}. Dysprosium ions have 7 f-electrons with J=15/2 for the single ion angular momenta in the ground state, with the crystal field levels all Kramers doublets; holmium ions, on the other hand, have 8 f-electrons with an integer J=8, so its crystal field spectrum has both singlet and doublet energy levels \cite{Tomasello}. The magnetic heat capacity of both materials has a broad peak at low temperatures (1.2K for Dy$_2$Ti$_2$O$_7$ and 1.9K for Ho$_2$Ti$_2$O$_7$) \cite{Ramirez,Lau,Bramwell2001}. However, Ho$_2$Ti$_2$O$_7$ has another peak in heat capacity at much lower temperatures, which is believed to be due to its active nuclear magnetism \cite{Bramwell2001,VanKempen}. Interpreting the magnetization dynamics of these two systems has proven challenging. Nearest neighbor spin-ice-based models for spin dynamics predict inelastic neutron scattering intensity patterns where ''pinch points'' are significantly smeared out compared to the experiments \cite{Fennell2007,Fennell2009}, and exchange interactions up to third nearest neighbor must be included in such models to replicate the full complexity of these neutron scattering results \cite{Yavorskii}. Models for the magnetic susceptibility $\chi(\omega,T)$ based on dipolar-spin-ice \cite{Takatsu}, or on the dynamics of a dilute gas of mobile monopoles \cite{Castelnovo2008} representing transitions from two-in-two-out to three-in-one-out configurations\cite{Ryzhkin,JaubertNatPhys} (see Fig. \ref{fig:Fig1} b), show significant deviations from the experimental data \cite{Takatsu}. Moreover, predictions for the temperature dependence of microscopic magnetic relaxation times $\tau$ considerably underestimate the divergence of $\tau$ at low temperatures \cite{Yaraskevitch}. Finally, in the absence of magnetic fields, no magnetic order has been detected in either compound \cite{Melko}. 

A recent proposal \cite{Kassner} that the magnetic state of Dy$_2$Ti$_2$O$_7$ is the magnetic analog of the diverging viscosity state found in glass-forming dipolar liquids \cite{Ediger,Tarjus,Cavagna,deSouza} provides a different perspective. Classical glass-forming liquids exhibit universally a super-Arrhenius divergence of microscopic dipolar relaxation times $\tau_0(T)$ of the Vogel-Tammann-Fulcher (VTF) form $\tau_0(T)=A\text{exp}(DT_0/(T-T_0 ))$ \cite{Bohmer}, a dielectric function $\epsilon (\omega,T)$ of the Havriliak-Negami (HN) form $\epsilon (\omega,T)= \epsilon_\infty+  \epsilon_0/(1+(i\omega\tau_{HN} )^\alpha )^\gamma$  \cite{HN,Havriliak}, and a related time-domain relaxation described by the Kohlrausch-Williams-Watts (KWW) form \cite{Kohlrausch} $\epsilon (t) = \epsilon_0 \text{exp} [-(t/\tau_{KWW})^\beta ]$. Observation of this combined VTF/HN/KWW phenomenology provides a strong clear identifier of a supercooled glass-forming dipolar fluid \cite{Ediger,Tarjus,Cavagna,deSouza}. Dy$_2$Ti$_2$O$_7$ was found to exhibit a precise HN form for its magnetic susceptibility $\chi (\omega, T)$, a general KWW form for the magnetic relaxation, and diverging microscopic magnetic relaxation rates on a VTF trajectory, implying that it is, by analogy, a glass-forming magnetic liquid \cite{Kassner}. Here we explore if magnetic fluids with such a phenomenology could be more general in the lanthanide pyrochlore magnetic materials. 

Even if glass-forming spin-liquid phenomenology were common to such materials, the microscopic parameters are still likely to be specific to each compound. In glass-forming dipolar liquids, the measure of the correlated dipole dynamics is called the fragility, D, \cite{Ediger2000} and it characterizes the degree of spatial heterogeneity. D is an indicator of the spread of microscopic relaxation times over different close-by regions in the liquid. The smaller the value of D, the more fragile the liquid and the more spatially heterogeneous its dynamics \cite{deSouza,Ediger2000}. By analogy, a more fragile glass-forming spin-liquid would mean an enhancement of the super-Arrhenius behavior of its magnetic relaxation times upon cooling. Such a situation could be caused by less efficient tunneling between spin configurations, due perhaps to differences in monopole creation energies and hopping rates. For the pyrochlore magnets discussed, the chemical potential for monopole-pair generation is dependent on nearest neighbor coupling, J$_{\text{eff}}$ \cite{JaubertHoldswort}, which is 1.1K for Dy$_2$Ti$_2$O$_7$ and 1.8K for Ho$_2$Ti$_2$O$_7$. Moreover, the theoretical rate of tunneling of the monopole excitations depends on the off-diagonal components of the dipolar interactions of neighboring spin \cite{Ehlers}. The reason for this is the strong Ising-like behavior of the magnetic ions \cite{Rau2015}, with the energy barriers to the first excited crystal field state being $\Delta \sim 240$K for Ho$_2$Ti$_2$O$_7$ and $\sim$ 380K for Dy$_2$Ti$_2$O$_7$ \cite{JaubertHoldswort}. Additionally, the fact that the effective energy scale for spin-flip dynamics is on the order of several J$_{\text{eff}}$ rather than $\Delta$ implies that spin flips occur by quantum tunneling \cite{Ehlers}. Since the transverse field effects in Ho$_2$Ti$_2$O$_7$ are more pronounced than in Dy$_2$Ti$_2$O$_7$, resulting in a more effective quantum tunneling at low magnetic fields \cite{Tomasello}, monopole hopping in Ho$_2$Ti$_2$O$_7$ is expected theoretically to be more efficient. In that case, one might anticipate a less fragile glass-forming spin-liquid in Ho$_2$Ti$_2$O$_7$ as compared to Dy$_2$Ti$_2$O$_7$.

	\section{Experimental setup} 

To explore the relationship between the magnetization dynamics of Ho$_2$Ti$_2$O$_7$ and Dy$_2$Ti$_2$O$_7$ we used a boundary-free arrangement to measure the AC susceptibility and time dependent magnetization relaxation characteristics of the two materials at T$\leq$2 K. Single crystals of these materials were grown as boules in O$_2$ gas under 2 atm pressure in an optical floating zone furnace \cite{Dabkowska}, and were subsequently cut into disks of diameter $\sim$ 6mm and thickness $\sim$ 1mm (see Fig. \ref{fig:Fig1} d). For the boundary-free magnetization measurements, holes of $\sim$ 2.5mm diameter were drilled through the center of disk shaped samples of both Ho$_2$Ti$_2$O$_7$ and Dy$_2$Ti$_2$O$_7$ crystals. A superconducting toroidal solenoid (STS) was then made by winding a 0.09mm diameter NbTi wire around the toroidal samples (Fig. \ref{fig:Fig1} c,d). Using a toroidal geometry for both the samples and the magnetization sensors means that the superconducting toroidal solenoid can be used to both drive magnetization flow azimuthally and to simultaneously and directly determine $dM/dt$ throughout. More importantly, it removes any boundaries in the direction of the magnetization transport (Fig. \ref{fig:Fig1} c). The coil EMF due to changes in both the applied azimuthal field $H(t)$ and sample magnetization $M(t,T)$ is given by

\begin{equation}
V_{total} (t,T)= -\mu_0 NA\biggl(\frac{dH(t)}{dt}+\frac{dM(t,T)}{dt}\biggr)
\end{equation}

where $N$ is the number of turns in the solenoid and $A$ is the effective cross-sectional area of the solenoid. Thus, the EMF due to magnetization dynamics in the sample is 

\begin{equation}
V(t,T)= -\mu_0 NA \frac{dM(t,T)}{dt}
\end{equation}

For an applied AC field %$H(t)=H_0 \text{exp}⁡(i\omega t)$ we expect that $M(t,T)=\text{M}(\omega ,T)\text{exp}(i\omega t)$ with some complex amplitude $\text{M}(\omega, T)$, so that

\begin{equation}
V(\omega,T)= -i\mu_0 NA \omega \text{M}(\omega,T)
\end{equation}

The definition of the magnetic susceptibility is

\begin{equation}
\text{M}(\omega,T)=\chi(\omega,T)H(\omega)
\end{equation}

In a solenoid $H_0=nI$, where $n$ is the number of turns per unit length, so the EMF is given by

\begin{multline}
V=V_x+iV_y=-i\mu_0 NA\omega \chi(\omega,T)H_0=\\ -i I\omega L [\chi' (\omega,T)-i\chi'' (\omega,T)]
\label{eq:Eq5}
\end{multline}

where $L$ is the geometric inductance of the STS pickup coil. Currents of up to 200 mA can be applied, using low temperature Nb crimp joints, to the STS coils, yielding azimuthal applied fields of magnitude up to $|$B$|$=2.5 mT or $|$H$|$=2200 A/m. Such fields are orders of magnitude smaller than those required
to flip spins in these compounds. In addition, this azimuthal field covers a wide range of crystallographic planes in its path.  
The AC susceptibility of the compounds measured was determined typically by applying $\sim$10 mA currents in a frequency range of 10 Hz – 100 kHz using a 4-probe impedance measurement of the STS. The inductance, L, of the STS was measured at T=50 mK, where neither of the materials show any magnetic activity in the frequency range measured, and then used in equation \ref{eq:Eq5} to calculate the susceptibility $\chi (\omega,T)$ data from the voltage readings.

	\section{Results}
\begin{figure*}[]
\includegraphics[width=1\textwidth]{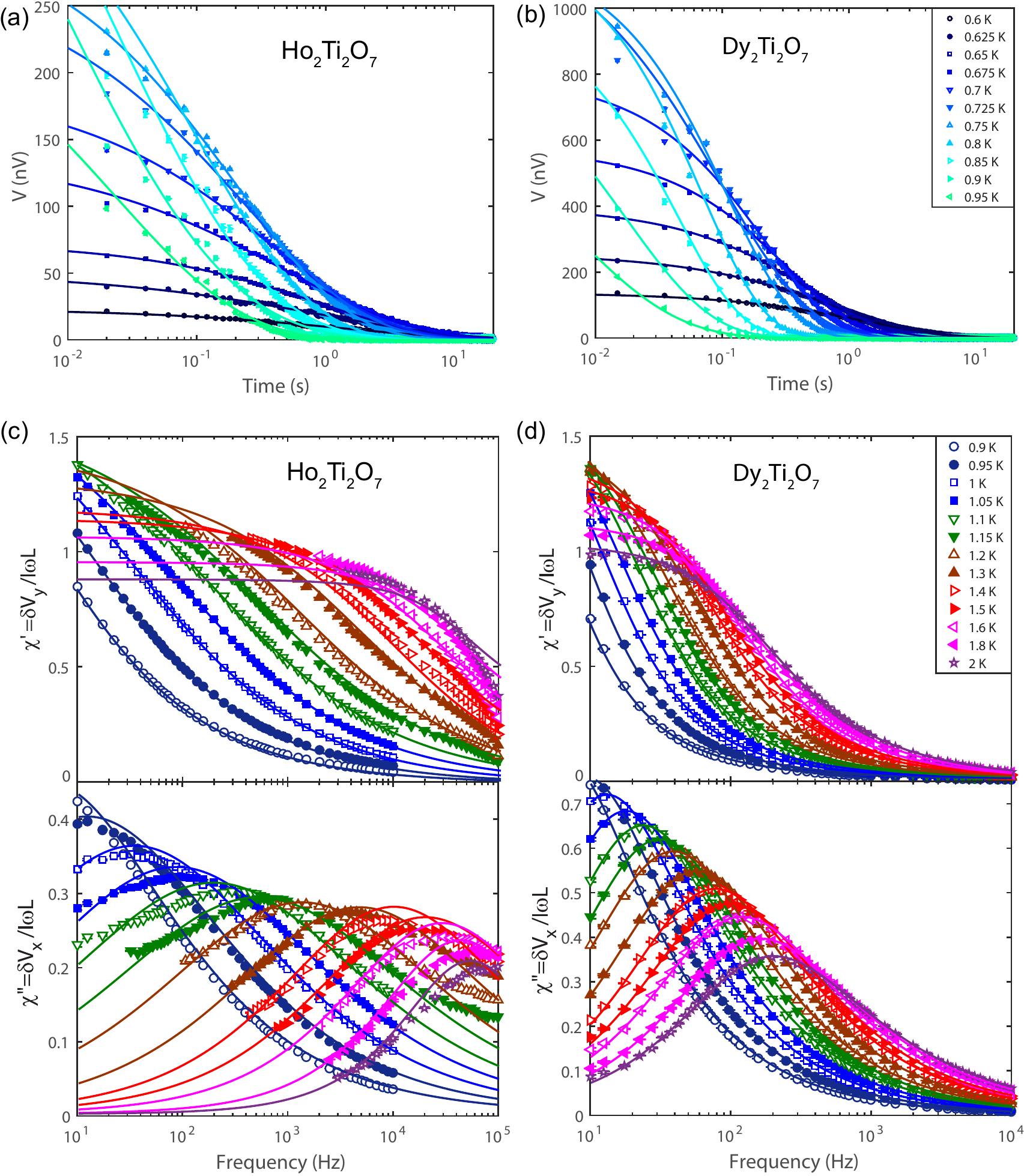} {}
\caption{
Comparison of the measured DC voltage and AC susceptibilities of the two lanthanide titanate pyrochlore materials at various temperatures. 
(a) and (b): Time domain magnetization relaxation measurements. At every temperature, the EMF across the STS is measured immediately after a step function change in the azimuthal magnetic field. This EMF is a measure of the magnetization change with time, $dM/dt$, in Ho$_2$Ti$_2$O$_7$ (a) and Dy$_2$Ti$_2$O$_7$ (b). Fine colored curves show the fit to the KWW functional form $V(t)=V_0 \text{exp⁡}(-(t/\tau)^\beta)$ at different temperatures for both materials. In both (a) and (b) the error bars are generally smaller than the symbols used to represent the data values. 
(c) Measured real (top panel) and imaginary (bottom panel) parts of $\chi(\omega,T)$ of Ho$_2$Ti$_2$O$_7$ and (d) Dy$_2$Ti$_2$O$_7$. The AC susceptibility was calculated from the four-probe measurement of the self-inductance of the STS according to equation \ref{eq:Eq5} in the text. Fine black lines associated with each set of symbols representing measured $\chi(\omega,T)$ at a given T are the Havriliak-Negami forms of the susceptibility fitted at that T. In both (c) and (d) all error bars are shown but are generally much smaller than the symbols used to represent the data values.
}
\label{fig:Fig2}
\end{figure*}

\begin{figure*}[]
\includegraphics[width=1\textwidth]{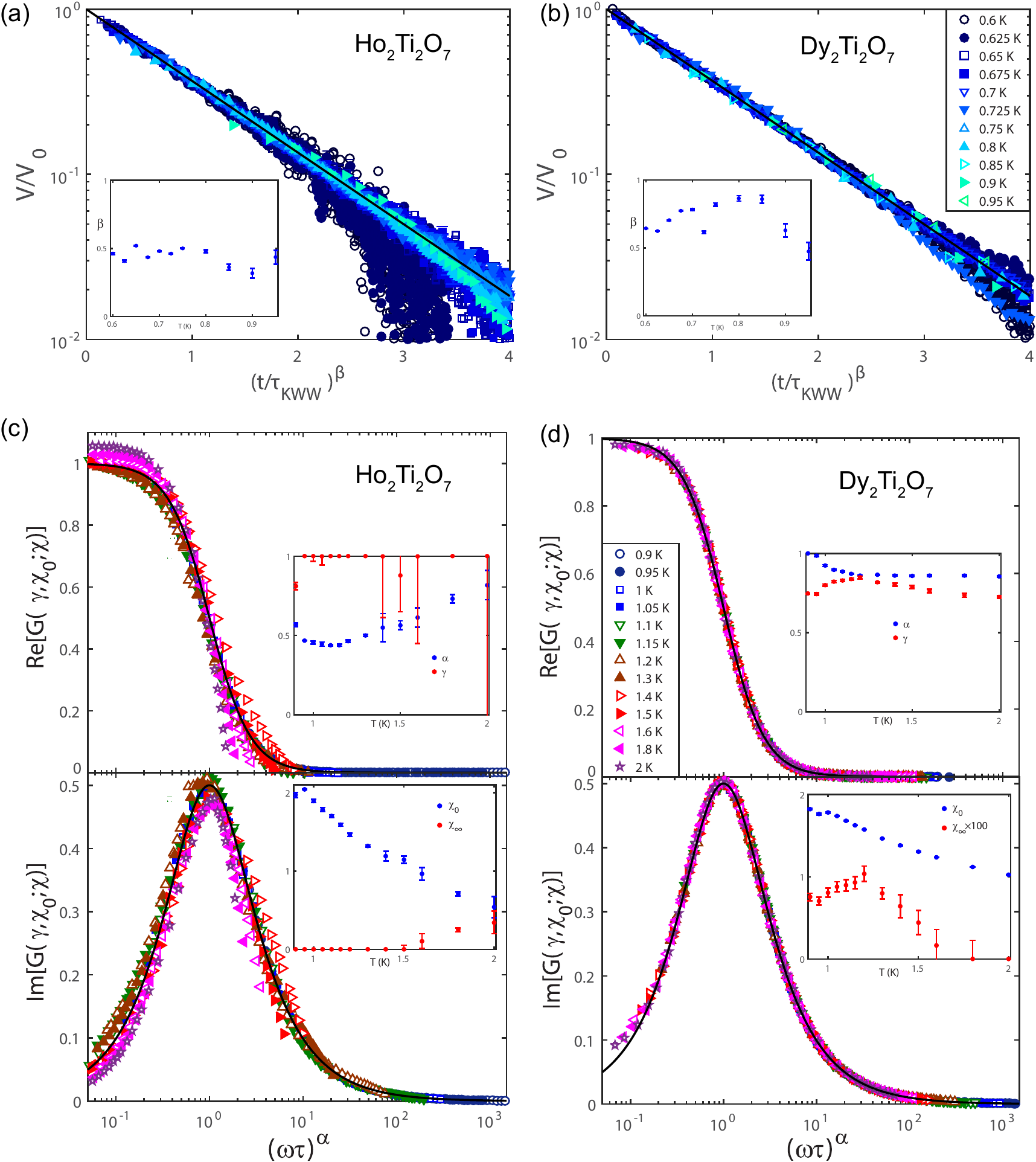} {}
\caption{
Comparison between Ho$_2$Ti$_2$O$_7$ and Dy$_2$Ti$_2$O$_7$ of the collapsed data sets of both DC and AC susceptibility at all measured temperatures, using the fit parameters from fitting the data to the KWW and HN forms (see text and Fig. \ref{fig:Fig2} c,d). 
(a) and (b) The horizontal axis is modified to account for the temperature dependence of the fit parameters of the time domain magnetization relaxation measurement, $\beta(T)$ and $\tau(T)$ for (a) Ho$_2$Ti$_2$O$_7$ and (b) Dy$_2$Ti$_2$O$_7$. The vertical axis is modified from raw data by dividing by the fit parameter $V_0$ of the KWW form. The collapse of the data for both Ho$_2$Ti$_2$O$_7$ and Dy$_2$Ti$_2$O$_7$ onto a single straight line shows a universal KWW form for all temperatures measured. Insets: temperature dependence of the exponent $\beta$ for Ho$_2$Ti$_2$O$_7$ (a) and for Dy$_2$Ti$_2$O$_7$ (b).
 (c) and (d) The real and imaginary parts of $\chi(\omega,T)$ for both Ho$_2$Ti$_2$O$_7$ and Dy$_2$Ti$_2$O$_7$, and the frequency, have been rescaled to account for the temperature dependence of all of the fit parameters in the HN form. The resulting scaled universal susceptibility function  $G(\gamma,\chi_0,\chi)$ (see text and Ref. \cite{Kassner}) for (c) Ho$_2$Ti$_2$O$_7$ and (d) Dy$_2$Ti$_2$O$_7$ from all $\omega$ and T converges onto a single curve for both materials, shown by the overlaid fine black curves. Insets: temperature dependence of the HN exponents from equation \ref{eq:Eq6} in the text, $\alpha$ and $\gamma$, for Ho$_2$Ti$_2$O$_7$ and Dy$_2$Ti$_2$O$_7$ (top panel of (c) and (d) respectively), and the temperature dependence of the coefficients, $\chi_0$ and $\chi_\infty$, for both materials (bottom panel of (c) and (d) respectively).}
\label{fig:Fig3}
\end{figure*}

During transient data acquisition, the voltage over the STS was measured every 20 ms throughout the following excitation protocol: (a) apply magnetic field in a clockwise direction by turning on a current I=50 mA in the STS, (b) set the field to zero by turning off the current, (c) apply a magnetic field in the counter clockwise direction by turning on a current I=-50 mA in the opposite direction and (d) again zero the field. This protocol was repeated 150 times per temperature for each material at each temperature, and the results were averaged to improve data quality and fitting. For both materials, no difference in relaxation characteristics was observed when the magnetic field was turned on or off, as well as when the magnetic field was applied in one azimuthal direction or the opposite. At long times, after the initial sharp change in the field, the EMF that was generated in the STS decayed to zero, indicating that $J=dM/dt$ \cite{Ryzhkin,Kassner} always decays to zero, despite the fact there are no terminating boundaries in our geometry. Figs. \ref{fig:Fig2} a,b depict the measured magnetization relaxation characteristics $dM/dt$ of Ho$_2$Ti$_2$O$_7$ and Dy$_2$Ti$_2$O$_7$ respectively in the temperature range 0.6 K - 0.95 K. The plots show the measured voltage induced across the STS by the magnetization dynamics of the sample versus time after the application of a DC field. These data sets at each temperature were fitted by a KWW type stretched exponential decay $V(t)=V_0 \text{exp}(-(t/\tau)^\beta)$, with fits shown in Fig. \ref{fig:Fig2} a,b as fine colored curves. Although a simple exponential decay cannot fit any of these data at any temperature, the KWW form provides an excellent fit for all. The insets of Fig.  \ref{fig:Fig3} a,b show how the stretching parameter, $\beta$, is different from unity over the temperature range of the DC measurements for both compounds. More importantly, Figs. \ref{fig:Fig3} a,b reveal the universal applicability of the KWW form of both samples for the whole temperature range. Here, the normalized EMF $V(t)/V_0$ is plotted against the modified time parameter $x=(t/\tau)^\beta$  for each temperature, with the result that all the magnetization transient data from both materials collapse onto a single line with unit slope. This remarkable agreement of magnetization decay dynamics of both Ho$_2$Ti$_2$O$_7$ and Dy$_2$Ti$_2$O$_7$ with a KWW form implies that both these systems are in the same state, a glass-forming spin-liquid.

The AC magnetic susceptibility of the toroidal samples of Ho$_2$Ti$_2$O$_7$  and Dy$_2$Ti$_2$O$_7$  was measured in the temperature range 0.9 K-2 K. For the compounds in this paper, we observed that below 0.5 K the EMF generated in the STS, in the frequency range reported, showed virtually no temperature dependence down to 50 mK, the lowest temperature at which AC measurements were attempted. The AC voltage measured at the lowest temperature was subtracted from the measurement at the temperatures of interest to deduce the susceptibility $\chi(\omega,T)=\chi'(\omega,T)-i\chi''(\omega,T)$. Figs. \ref{fig:Fig2} c,d present the measured real ($\chi '$) and imaginary ($\chi ''$) parts of the susceptibilities for the samples measured versus frequency, in the range 10-$10^5$ Hz for Ho$_2$Ti$_2$O$_7$ and 10-$10^4$ Hz for Dy$_2$Ti$_2$O$_7$. The data sets taken at different temperatures are labeled by a color/symbol code as indicated. 

Models of AC susceptibility that assume a single relaxation time of the Debye form $\chi'-i\chi ''=\chi_0/([1+(i\omega\tau)])$, for example those of free monopole motion \cite{Takatsu}, are not compatible with the measured $\chi(\omega,T)$ for either Ho$_2$Ti$_2$O$_7$ or Dy$_2$Ti$_2$O$_7$ at any frequency or temperature within these ranges (see Fig. \ref{fig:Fig2}). By contrast, a Havriliak-Negami form modifies the simple Debye susceptibility with two exponents, $\alpha$ and $\gamma$, and corresponds to a system where there is a distribution of relaxation times

\begin{equation}
\chi'-i\chi''=\frac{\chi_0}{[1+(i\omega\tau)^\alpha ]^\gamma }+\chi_\infty
\label{eq:Eq6}
\end{equation}

Figs. \ref{fig:Fig2} c,d depict our measured data for both Ho$_2$Ti$_2$O$_7$ and Dy$_2$Ti$_2$O$_7$ respectively. The top panels show the real part of the measured magnetic susceptibility versus frequency, as calculated from our measured voltage, using equation \ref{eq:Eq5}, and the bottom panels present the imaginary part of the susceptibility versus frequency. The different colors/symbols show data from different temperatures in the range 0.9K to 2K. For Dy$_2$Ti$_2$O$_7$, both exponents of the HN fit (equation \ref{eq:Eq6}) deviate from unity for the majority of the temperature range (inset of Fig. \ref{fig:Fig3} d), whereas for Ho$_2$Ti$_2$O$_7$ $\gamma$ is around unity for most temperatures within error. Overall, the susceptibility for both materials shows a very good global agreement with the HN form for all temperatures and frequencies measured, as demonstrated by the fine lines in Fig. \ref{fig:Fig2} c,d. Fig. \ref{fig:Fig3} c,d show the collapse of these dynamical susceptibility $\chi(\omega,T)$ data for all temperatures and both materials onto the single HN form \cite{Kassner} as indicated by the fine solid curves. The horizontal axis in the figure is the frequency, scaled by the HN parameters, and the vertical axes are the real and imaginary parts of the scaled HN susceptibility $G(\gamma,\chi_0,\chi)$ (a full mathematical derivation can be found in Ref. 
\cite{Kassner}).

%\begin{multiline} 
\begin{equation}
\begin{split}
G(\gamma,\chi_0,\chi)=
\biggl(\frac{\chi'^2+\chi''^2}{\chi_0^2}\biggr)^{\frac{1}{2\gamma}}\biggl(\text{cos}\biggl(\frac{1}{\gamma}\text{arctan}\frac{\chi''}{\chi'}\biggr)\\
-i \text{sin}\biggl(\frac{1}{\gamma}\text{arctan}\frac{\chi''}{\chi'}\biggr)\biggr)
\end{split}
\label{eq:Eq7}
\end{equation}
%\end{multiline}

The scaling parameters, which are the fit parameters of the HN form for each temperature, are plotted in the insets of the figure. The quality of fits, while comprehensively good versus $\omega$ and $T$ for both materials (Fig. \ref{fig:Fig2} c,d), is obviously slightly different between Ho$_2$Ti$_2$O$_7$ and Dy$_2$Ti$_2$O$_7$ (Fig. \ref{fig:Fig3} c,d). This may not be surprising since the frequency-width and frequency-range of the data from the Ho$_2$Ti$_2$O$_7$ measurements is at least two orders of magnitude wider than that for Dy$_2$Ti$_2$O$_7$ (see Fig. \ref{fig:Fig2} c,d). In any case, this observation of a universal Havriliak-Negami form for all the $\chi(\omega,T)$ susceptibility data (Fig. \ref{fig:Fig3} c,d) constitutes a second robust indication that both these materials are homologous glass-forming spin-liquids.

\begin{figure*}[]
\includegraphics[width=1\textwidth]{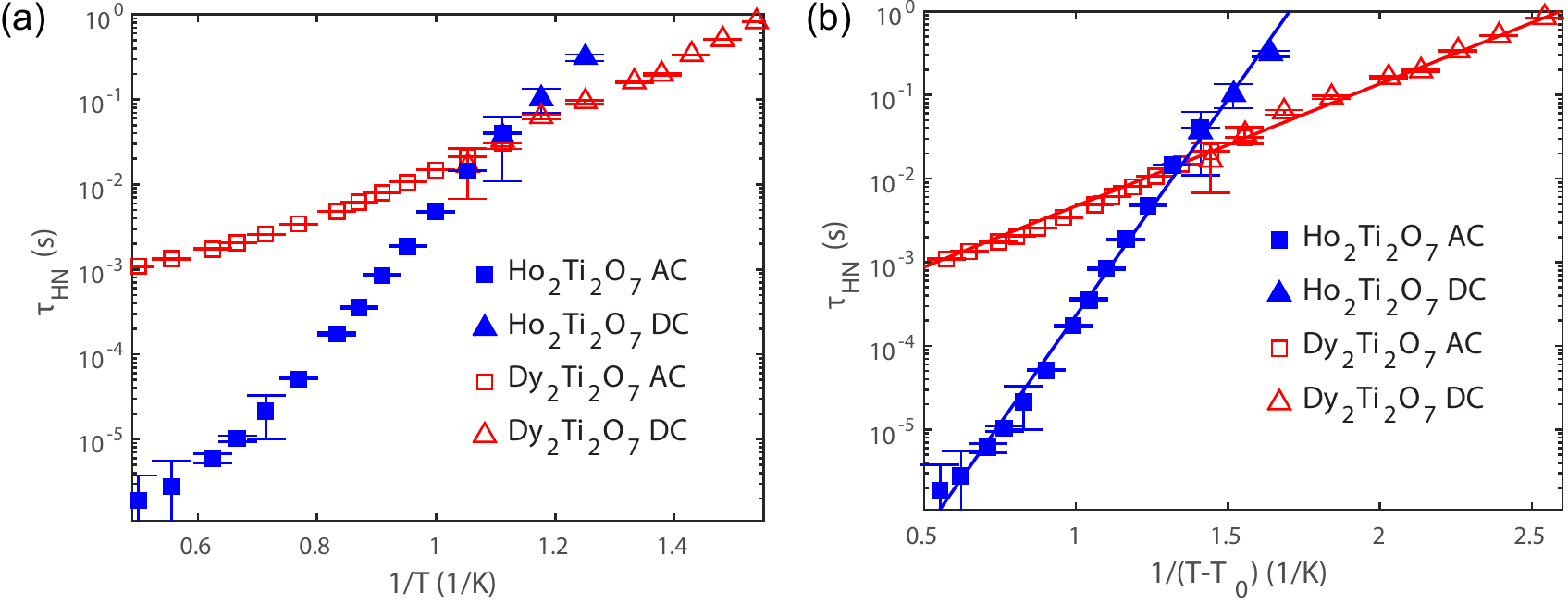} {}
\caption{
(a) Temperature dependence of the relaxation times obtained from the frequency (triangles) and time domain (squares) measurements for both Ho$_2$Ti$_2$O$_7$ and Dy$_2$Ti$_2$O$_7$. In the region where the relaxation times from both techniques overlap, they merge smoothly. The straight lines present the best fit to an Arrhenius form. (b) A super-Arrhenius VTF function of the form $\tau(T)=A\text{exp}(DT_0/(T-T_0 ))$ fits the temperature dependence of the relaxation times much better, yielding a divergence around $T_0\sim$200mK for both materials. The fragility indices, D, however, are quite different, with Dy$_2$Ti$_2$O$_7$ having a more fragile, D$\approx$14, hence heterogeneous, dynamics, as expected by the more efficient tunneling of monopoles in Ho$_2$Ti$_2$O$_7$ for which D$\approx$60 from the fit. Note that the x axis in (b) is slightly different for the two materials. 
 }
\label{fig:Fig4}
\end{figure*}

Finally, to explore the microscopic magnetic relaxation dynamics of these systems, we need a form to relate the relaxation times obtained from the time-domain measurements to those from the frequency-domain. Numerical studies have linked the exponents and relaxation time parameters of the two forms \cite{Alvarez}, which can be used for a unified analysis of our data. Using the values in Table I of Ref. \cite{Alvarez} we can generate values for $ \frac{\tau_{HN}}{\tau_{KWW}}$, enabling a conversion of the relaxation times from the time-domain measurements to the frequency-domain \cite{sup2}. 
Fig. \ref{fig:Fig4} a depicts the combined time- and frequency-domain relaxation-time data for Ho$_2$Ti$_2$O$_7$ (T$\geq$0.8 K) and Dy$_2$Ti$_2$O$_7$ respectively (the relaxation-times obtained from the time-domain measurement, $\tau_{KWW}$, were converted to $\tau_{HN}$ by the procedure described above) with the horizontal axis being the inverse temperature. Obviously, the relaxation-time data for both materials diverge on a trajectory that is faster than Arrhenius, which would produce a straight line in Fig. \ref{fig:Fig4} a. Indeed, many groups have previously reported relaxation-time data showing the general behavior of a divergence that is faster than Arrhenius \cite{Yaraskevitch,SnyderPRB,Quilliam,Matsuhira}, and in particular Ho$_2$Ti$_2$O$_7$ showing a stronger divergence.

However, when the temperature dependence of the relaxation-time is fit to a VTF form,  $\tau(T)=A\text{exp}\frac{DT_0}{T-T_0}$, as shown in Fig. \ref{fig:Fig4} b, our findings indicate that both Dy$_2$Ti$_2$O$_7$ and Ho$_2$Ti$_2$O$_7$ exhibit non-Arrhenius slowing. This form yields A$\approx8\times10^{-10}$s, a fragility parameter D$\approx$60 and a VTF temperature $T_0\approx$191 mK for Ho$_2$Ti$_2$O$_7$, and A$\approx1.4\times10^{-4}$s, D$\approx$14 and T$_0\approx$257mK for Dy$_2$Ti$_2$O$_7$, signifying that Dy$_2$Ti$_2$O$_7$ is a more fragile spin-liquid. These specific parameters, resulting from the best fit to the VTF form give standard errors of 1 to 2 percent, and R$^2$ of 0.995 and 0.998 respectively.
Systematic errors arising from the fit procedure can be up to several tens of percent, but those have no impact on the resulting function, as can be clearly seen in Fig. \ref{fig:Fig4}b and by the high R$^2$. All of the data presented in figures \ref{fig:Fig2}, \ref{fig:Fig3} and \ref{fig:Fig4} are newly acquired for the purposed of comparison between the two compounds: the parameters of Dy$_2$Ti$_2$O$_7$ from this work agree well with previous work \cite{Kassner}. The fragility parameters and high-temperature relaxation-times depend strongly on the materials studied, but their T$_0$, the lowest temperature at which both materials may be expected, by analogy with glass-forming fluids, to enter a magnetic glass phase, are close in value. Overall, we find in the common VTF form for $\tau(T)$ (see Fig. \ref{fig:Fig4}b), a clear indication that both Dy$_2$Ti$_2$O$_7$ and Ho$_2$Ti$_2$O$_7$ are glass-forming spin-liquids.

	\section{Discussion}

Previous comparison studies of AC susceptibility of these two compounds \cite{Yaraskevitch,Quilliam} identified differences in the spread of their microscopic relaxation times. Specifically, the broadness of the absorption spectra, inferred from the width of the imaginary part of the magnetic susceptibility $\chi''(\omega,T)$ as function of frequency, was found to be greater for Ho$_2$Ti$_2$O$_7$ than for Dy$_2$Ti$_2$O$_7$. The spread of characteristic relaxation times $\tau$, as well as the asymmetry in $\chi''(\omega,T)$, were also found to be broader in Ho$_2$Ti$_2$O$_7$ \cite{Yaraskevitch}. The qualitative agreement of these works with the profile of the scaled susceptibilities $G(\gamma,\chi_0,\chi)$ shown in Fig. \ref{fig:Fig3} c,d, alongside the difference in the microscopic energy scales of Ho$_2$Ti$_2$O$_7$ and Dy$_2$Ti$_2$O$_7$ \cite{JaubertHoldswort} indicate that the differences between the two compounds are unlikely due to random disorder or off-stochiometry. In addition, previously reported relaxation times for Ho$_2$Ti$_2$O$_7$ and Dy$_2$Ti$_2$O$_7$, using the inverse of the angular frequency of the peak in $\chi''(\omega,T)$ (see Fig. 3 in ref. \cite{Yaraskevitch} for example and Ref. \cite{sup1}), exhibit the distinct characteristics shown in Fig. \ref{fig:Fig4} - the slope of Ho$_2$Ti$_2$O$_7$ is greater than that of Dy$_2$Ti$_2$O$_7$ and the relaxation times cross around 0.9 K.

The apparent inferiority of the functional HN fits to Ho$_2$Ti$_2$O$_7$ as compared to Dy$_2$Ti$_2$O$_7$ is probably due to a combination of the more than two orders of magnitude wider spread of the relaxation times, and smaller signal sizes. However, these are still the best internally consistent analytic forms for $\chi(\omega,T)$ in these materials, consistent for both the magnetization and susceptibility measurements and their resulting relaxation time temperature dependence. They yield parameters that agree both with prior works and do not contradict expectations from the different energy scales in these materials. 

The nomenclature of the proposed glass-forming spin-liquid (GFSL) motivates comparisons to existing spin glasses. Even though both have connection to glassy behavior, i.e. magnetic dynamics slow down with decrease in temperature, these two classes of materials are quite physically distinct. The difference between spin glasses and the proposed glass-forming spin-liquid state is both conceptual and in the details. First, quenched disorder is key for spin glasses, whereas there is no intrinsic disorder in the spin-ice compounds we study. Secondly, one of the clear signatures of spin glasses is the presence of a sharp cusp in the real part of the magnetic susceptibility at the transition temperature \cite{Mydosh}, whereas the magnetic susceptibility of GFSL has a very smooth profile. Lastly, one can distinguish between the two by measuring the magnetic noise spectrum. The noise spectrum of a system carries information about its fluctuations on a microscopic scale. In the spin glass state, for example, one would expect a 1/f noise to  present \cite{Bouchiat}.

The GFSL formalism, underpinning our studies, presents several verifiable predictions for spin-ices: the exact functional forms of the magnetization decay and ac susceptibility, and the divergence of the relaxation times. No other model can account for all of the observed behavior in these compounds in such great detail. In addition, the GFSL suggests a link to the underlying mechanisms through its fragility parameter. The fragility of conventional glass forming liquids reveals the degree of deviation of a system's relaxation time from the Arrhenius profile. The more fragile the glass former, the more curved its relaxation time divergence with respect to temperature. In the VTF formalism, this fragility is manifest as the slope of log($\tau$) vs 1/T i.e. the temperature dependent energy scale in the Arrhenius form. Therefore, the more fragile a material, the smaller the slope. Examination of the equation that describes the super-Arrhenius form, 
$\tau(T)=\tau_0\text{exp}\frac{DT_0}{T-T_0}$, shows that D is connected to the ac susceptibility characteristics of the glass former through the peak frequency, that is proportional to 1/$\tau$.

In the frustrated pyrochlore magnets we study, fragility appears to indicate the heterogeneous nature of the energy landscape in these materials, similar to what is found in glass-forming dipolar-liquids. Therefore, the difference in fragilities between the two materials may eventually be revealed as due to differences in the underlying correlated and frustrated microscopic behavior of the magnetic monopoles, that is hypothesized to cause the magnetic relaxation in these materials. Here, both monopole creation energies and hopping rates, as well as constraining by Dirac strings would influence the dynamics, and hence the energy landscape. The higher fragility parameter of Ho$_2$Ti$_2$O$_7$, indicating a less fragile GFSL, for example, is consistent with the expected efficient monopole hopping in Ho$_2$Ti$_2$O$_7$ \cite{Tomasello}. 

Finally, it seems plausible that the observed magnetic dynamics in Ho$_2$Ti$_2$O$_7$ and Dy$_2$Ti$_2$O$_7$ can be reconciled with the microscopic theory of emergent magnetic monopoles in these materials \cite{Castelnovo2008,JaubertHoldswort} by considering correlated transport of these quasiparticles. Here, flips of the real magnetic dipoles are recast as two opposite magnetic charges that, through a sequence of spin flips, may form a fluid of delocalized magnetic monopoles and anti-monopoles \cite{Castelnovo2008}. At low temperatures, these monopoles should form a dilute neutral gas whose transport characteristics control the magnetization dynamics and the susceptibility \cite{CastelnovoDH}. However, these monopoles are constrained by a network of Dirac strings, i.e. a trail of flipped spins left behind by a monopole traversing the interconnected tetrahedra. The non-Arrhenius sharp slowing down of spin dynamics may then be attributed to relaxation of Dirac strings \cite{JaubertHoldswort}.

	\section{Conclusion}

To summarize: for purposes of comparison, we measured the AC susceptibility $\chi(\omega,T)$ and time dependent magnetic relaxation behavior in the low-temperature magnetic states of the two materials Dy$_2$Ti$_2$O$_7$ and Ho$_2$Ti$_2$O$_7$. We used identical boundary-free sample geometries within a superconducting toroidal solenoid. 
We find that, for both materials, the DC relaxation follows a stretched exponential, KWW form (Fig. \ref{fig:Fig3} a,b), the AC susceptibility follows a HN form (Fig. \ref{fig:Fig3} c,d), and above 0.8 K the relaxation time for both materials diverges along a super-Arrhenius trajectory (Fig. \ref{fig:Fig4}). These phenomena all indicate that the magnetic state of these two distinct materials is the magnetic analog of a glass-forming dipolar liquid, which seems to be a previously unidentified characteristic of this class of frustrated magnetic materials.They were not anticipated by but appear to be consistent with the DSIM.
The differences between the parameters of this general glass-forming spin-liquid phenomenology for the two materials can offer an insight into the microscopic behavior generating these phenomena. 
Indeed, recent theoretical studies using the spin-ice Hamiltonian extended to include stronger next nearest neighbor interactions, do report the existence of new forms of dynamical magnetic heterogeneity with extremely slow relaxation times for some spins \cite{Rau,Udagawa}. Thus, the type of glass-forming spin-liquid phenomenology that we observe in Dy$_2$Ti$_2$O$_7$ and Ho$_2$Ti$_2$O$_7$, can exist, in theory, in dipolar spin-ice.

	\begin{acknowledgements}
We acknowledge useful and encouraging discussions with S. Bramwell, C. Castelnovo, J. Chalker, H. Changlani, M. Gingras, E.-A. Kim, M.J. Lawler and J. Sethna. J.C.S.D., R.D. and A. E. acknowledge support from the Moore Foundation's EPiQS Initiative through Grant GBMF4544. 
 \\
\end{acknowledgements}

%Author Information: anna.eyal@gmail.com

\bibliographystyle{spphys}       % APS-like style for physics
\bibliography{DHTO}

\end{document}